\documentclass{llncs}
\usepackage{graphicx}
\usepackage{amsmath}
\usepackage{xspace}
\usepackage{listings}
\usepackage{xcolor}
\usepackage{subcaption}
\usepackage{changebar}
\usepackage{marvosym}
\usepackage{url}

\newcommand{\ASP}{\textsf{Asp}\xspace}
\newcommand{\Asp}{\ASP}

\newcommand{\out}[1]{\overline{#1}}

\newcommand{\becomes}{\leftarrow}

\newcommand{\TT}{\mathsf{true}}

\newcommand{\lAnd}{\,\wedge\,}
\newcommand{\lOr}{\,\vee\,}

\newcommand{\tDia}[1]{\langle{#1}\rangle}
\newcommand{\tBox}[1]{[{#1}]}
\newcommand{\Player}{\mathsf{Player}}
\newcommand{\Opponent}{\mathsf{Opponent}}

\newcommand{\tA}{\mathsf{A}}
\newcommand{\tE}{\mathsf{E}}
\newcommand{\tG}{\mathsf{G}}
\newcommand{\tF}{\mathsf{F}}

\newcommand{\Atime}{\mathsf{time}}
\lstdefinelanguage{Asp}{}
\lstdefinestyle{aspstyle}{
  language = Asp, 
  basicstyle=\small\ttfamily,
  keywordstyle=\color{violet}\bfseries,
  morekeywords={contract, msg, var, ghost, const, state, when, by, where},
  morekeywords={bool,int,nat,map,seq,coin,token},
  morekeywords={initial,final},
  morekeywords={owner,creator,timer, address},
  morekeywords={log},   
  morekeywords={issues},   
  morekeywords={invariance,always,atstate,notby},
  morekeywords={reachability,goal,invariant,rank,witness},
  morecomment=[l]{//},
  morecomment=[s]{/*}{*/},
  commentstyle=\color{gray},
  numberstyle=\tiny\color{gray},
  numbers=left,
    numbersep=5pt,  
    stepnumber=1,
    showstringspaces=false,
    tabsize=1,
    breaklines=true,
    breakatwhitespace=false,
    xleftmargin=1.2em        % push numbering within boundaries
}

\definecolor{verylightgray}{rgb}{.97,.97,.97}

\lstdefinelanguage{Solidity}{
	keywords=[1]{anonymous, assembly, assert, balance, break, call, callcode, case, catch, class, constant, continue, constructor, contract, debugger, default, delegatecall, delete, do, else, emit, event, experimental, export, external, false, finally, for, function, gas, if, implements, import, in, indexed, instanceof, interface, internal, is, length, library, log0, log1, log2, log3, log4, memory, modifier, new, payable, pragma, private, protected, public, pure, push, require, return, returns, revert, selfdestruct, send, solidity, storage, struct, suicide, super, switch, then, this, throw, transfer, true, try, typeof, using, value, view, while, with, addmod, ecrecover, keccak256, mulmod, ripemd160, sha256, sha3}, % generic keywords including crypto operations
	keywordstyle=[1]\color{blue}\bfseries,
	keywords=[2]{address, bool, byte, bytes, bytes1, bytes2, bytes3, bytes4, bytes5, bytes6, bytes7, bytes8, bytes9, bytes10, bytes11, bytes12, bytes13, bytes14, bytes15, bytes16, bytes17, bytes18, bytes19, bytes20, bytes21, bytes22, bytes23, bytes24, bytes25, bytes26, bytes27, bytes28, bytes29, bytes30, bytes31, bytes32, enum, int, int8, int16, int24, int32, int40, int48, int56, int64, int72, int80, int88, int96, int104, int112, int120, int128, int136, int144, int152, int160, int168, int176, int184, int192, int200, int208, int216, int224, int232, int240, int248, int256, mapping, string, uint, uint8, uint16, uint24, uint32, uint40, uint48, uint56, uint64, uint72, uint80, uint88, uint96, uint104, uint112, uint120, uint128, uint136, uint144, uint152, uint160, uint168, uint176, uint184, uint192, uint200, uint208, uint216, uint224, uint232, uint240, uint248, uint256, var, void, ether, finney, szabo, wei, days, hours, minutes, seconds, weeks, years},	% types; money and time units
	keywordstyle=[2]\color{teal}\bfseries,
	keywords=[3]{block, blockhash, coinbase, difficulty, gaslimit, number, timestamp, msg, data, gas, sender, sig, value, now, tx, gasprice, origin},	% environment variables
	keywordstyle=[3]\color{violet}\bfseries,
	identifierstyle=\color{black},
	sensitive=true,
	comment=[l]{//},
	morecomment=[s]{/*}{*/},
	commentstyle=\color{gray}\ttfamily,
	stringstyle=\color{red}\ttfamily,
	morestring=[b]',
	morestring=[b]"
}

\lstdefinestyle{soliditystyle}{
	language=Solidity,
	extendedchars=true,
	basicstyle=\footnotesize\ttfamily,
	showstringspaces=false,
	showspaces=false,
	tabsize=2,
	breaklines=true,
	showtabs=false,
	captionpos=b
}

\lstdefinelanguage{Viper}{
  keywords=[1]{method, returns, requires, ensures, var},
  keywordstyle=[1]\color{purple}\bfseries,
  keywords=[2]{Address, Coin, Int, Bool, Seq, Map, Tuple},
  keywordstyle=[2]\color{teal}\bfseries,
  comment=[l]{//},
  morecomment=[s]{/*}{*/},
  commentstyle=\color{gray}
}
\lstdefinestyle{viperstyle}{
  language = Viper, 
	extendedchars=true,
	basicstyle=\footnotesize\ttfamily,
	showstringspaces=false,
	showspaces=false,
	tabsize=2,
	breaklines=true,
	showtabs=false,
	captionpos=b
}

\lstdefinestyle{smallaspstyle}{
  language = Asp, 
  basicstyle=\footnotesize\ttfamily,
  keywordstyle=\color{black}\bfseries,
  morekeywords={contract, msg, var, const, state, when, by},
  morekeywords={bool,int,nat,map,seq,coin,token},
  morekeywords={initial,final},
  morekeywords={owner,creator,timer},
  morekeywords={log},
  morekeywords={issues},              
  morecomment=[l]{//},
  morecomment=[s]{/*}{*/},
  commentstyle=\color{gray}
}

\lstdefinestyle{smallviperstyle}{
  language = Viper, 
  basicstyle=\footnotesize\ttfamily,
  keywordstyle=\color{black}\bfseries,
  morekeywords={method, returns, requires, ensures},
  morekeywords={Address, Coin, Int, Bool, Seq, Map, Tuple},
  morekeywords={var},
  morecomment=[l]{//},
  morecomment=[s]{/*}{*/},
  commentstyle=\color{gray}
}

\newcommand{\aspinline}[1]{\lstinline[style=aspstyle]{#1}}

\newcommand{\elide}[1]{}

\begin{document}

\title{Constructing Trustworthy Smart Contracts}

\author{Devora Chait-Roth\inst{1}
\and
Kedar S. Namjoshi\inst{2}\textsuperscript{(\Letter)}}

\institute{
 New York University, New York, NY, USA \email{dc4451@nyu.edu}
 \and
 Nokia Bell Labs, Murray Hill, NJ, USA \email{kedar.namjoshi@nokia-bell-labs.com}
}

\maketitle

\begin{abstract}
  Smart contracts form the core of Web3 applications. Contracts mediate the transfer of cryptocurrency, making them irresistible targets for hackers. We introduce \ASP, a system aimed at easing the construction of provably secure contracts. The \Asp system consists of three closely-linked components: a programming language, a defensive compiler, and a proof checker. The language semantics guarantee that  \Asp contracts are free of commonly exploited vulnerabilities such as arithmetic overflow and reentrancy. 
The defensive compiler enforces the semantics and translates \Asp to Solidity, the most popular contract language.
Deductive proofs establish functional correctness and freedom from critical vulnerabilities such as unauthorized access.

\end{abstract}

\section{Introduction}
\label{sec:introduction}

Decentralized blockchain-based systems such as Ethereum have ushered in a new computation paradigm dubbed “Web3.'' Smart contracts form the core of Web3.  
A smart contract (``contract'' for short) is a program whose code and execution are recorded on a blockchain. A contract facilitates an exchange of value, typically cryptocurrency, between contract participants.

Trust is a central concern for Web3 applications, as there is no central trusted authority and the participants need not have prior trust relationships. 
Blockchain properties play a crucial role in building trust in contract execution. Every contract is stored on the blockchain, ensuring that its code is open and immutable. Every contract transaction is executed in a replicated manner, guarding against the possibility of machine failures or compromised execution engines.

Although these mechanisms provide a solid foundation, contracts are but programs, and programs may have errors. Immutability of code and fault-tolerant execution are of no help if contract code is buggy and vulnerable to attack. As contracts handle large amounts of cryptocurrency, they are irresistible targets for hackers, who stole nearly $\$4$ billion in 2022 and \$2 billion in 2023 from buggy contracts~\cite{cnbc-2022,certik-2023}. In response to this threat, developers audit contracts through standard processes: testing, static analysis, and expert code review. While these methods uncover some bugs, they are inadequate at guaranteeing security. 

As a result, \cite{verx,smartpulse,DBLP:journals/pacmpl/BramEMSS21} and others have developed  \emph{formal verification} methods for contracts. However, formally verifying contracts written in  standard languages such as Solidity or Rust is difficult, in part due to complex language constructs and the need to reason about reentrancy.\footnote{`Reentrancy' is an execution pattern where a malicious contract forces an internal contract-invocation loop that drains cryptocurrency from the target contract.} Non-standard contract languages have emerged in response, but have drawbacks. FSolidM~\cite{DBLP:conf/fc/MavridouL18,DBLP:conf/fc/MavridouLSD19} models contracts as finite state machines extended with Solidity actions, but verification is limited to propositional CTL properties. Scilla~\cite{DBLP:journals/pacmpl/SergeyNJ0TH19} compiles contracts to Coq for analysis, but proofs in Coq require substantial expertise and effort. Move~\cite{move-whitepaper-2019} has impressive automated verification support, but Move contracts cannot execute on common blockchains such as Ethereum.

\begin{figure}
\begin{center}
\includegraphics[scale=.6]{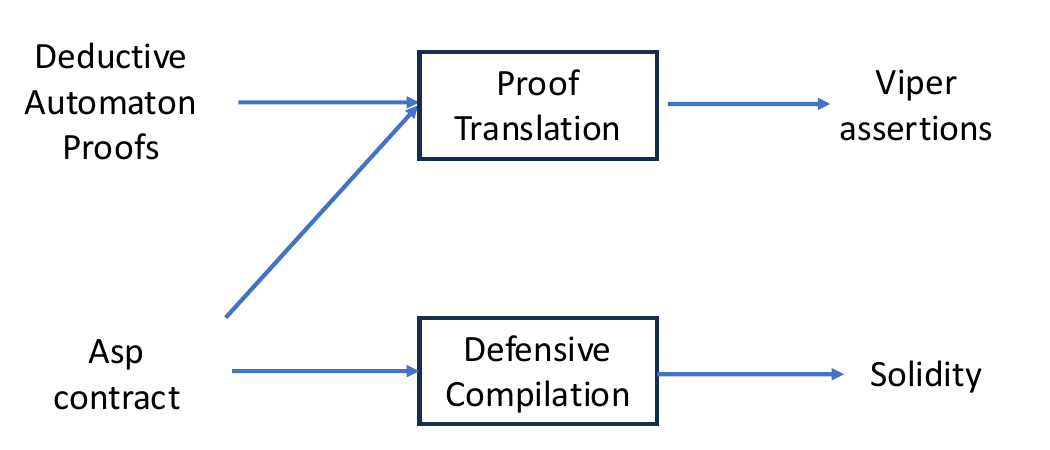}
\caption{The \Asp pipeline. The Viper system is used to verify proof assertions.} 
\label{fig:pipeline}
\end{center}
\end{figure}

This paper introduces the \Asp system, which is aimed at easing the construction of provably secure contracts while addressing these drawbacks. It has three tightly-linked components, as shown in Figure~\ref{fig:pipeline}: the \Asp contract language, a defensive compiler, and a deductive proof checker. 
The \Asp language provides abstractions that simplify verification; 
the compiler translates \Asp contracts to run on standard blockchains; and the \Asp proof checker is used to verify safety and liveness properties that users state directly on the \Asp contracts. 

\Asp combines well-known notions (such as state machines and deductive proofs) but differs from prior work on contract verification in its emphasis on abstraction coupled with defensive compilation. We illustrate the \Asp language and particularly its  abstraction mechanisms through a variety of examples, show how the use of abstractions simplifies proofs, and describe the implementation of the compiler and proof checker. Our \Asp prototype is implemented in about 3000 lines of OCaml and 100 lines of Viper; examples of \Asp contracts (with proofs) are available at \url{https://github.com/DebraChait/Asp-example-contracts}.
We summarize the three components of the \Asp system next.

\paragraph{The \Asp Language}
A programming language typically strikes a balance between ease of expression and ease of analysis. Most contracts today are written in the languages Solidity, Vyper, Ink! and Rust. These are full-fledged programming languages, which complicates verification. 
With \Asp, we aim to achieve a good balance through a programming model that employs \emph{abstractions}. 
Coins, tokens, timers, and addresses -- Web3 abstractions ordinarily represented at a low level -- are provided as abstract types in \Asp.
Abstractions ease programming and simplify analysis, as the abstract operations are few in number and have precise semantics. 
The \Asp language semantics inherently forbids arithmetic overflow, out-of-bounds accesses, and reentrancy, eliminating those common vulnerabilities as concerns for a contract programmer.

An \Asp contract has a finite-state machine skeleton that is augmented with actions on state variables defined over abstract data types. This structure naturally models real-world contract execution.
For instance, consider sending a package through the mail--a contract between the sender and the post office. This contract passes through the stages of preparation, payment, transit, and delivery; these stages are naturally modeled by a finite-state machine.  In conventional languages such as Solidity, such structure must be encoded and enforced implicitly,  obscuring it and complicating analysis.

\Asp contracts interact through synchronized message transfers. This also models the event-driven structure of real-world contracts: for instance, the change from the transit to the delivery stage in the prior example is through the event of delivering the package. 

\paragraph{The \Asp Defensive Compiler}
The \Asp compiler translates contracts to Solidity, the most popular contract language. Compilation preserves \Asp semantics, implements the high-level abstractions, and supports the message-transfer view of communication. The compiler adds auxiliary \emph{defensive} code to enforce the language semantics and check properties dynamically that are difficult to prove statically. Development of a compiler back-end for Ink! is in progress; compilation to other contract languages follows the same design. Compilation assures the portability of verified \Asp contracts to a variety of blockchains. 

\paragraph{The \Asp Proof Checker}
Although the \Asp semantics eliminates many common vulnerabilities, it cannot rule out all of them.
Thus, one must prove that contract behavior does not, for instance, allow unauthorized access to stored cryptocurrency, or that the contract cannot be placed in a `frozen' state. (These are commonly exploited vulnerabilities.\footnote{\url{https://info.merklescience.com/april-2023-hackhub-report}}) The first is a safety property, which \Asp users can establish through a standard automaton-based proof system. \Asp allows auxiliary `ghost' state to be added to a contract purely for the purposes of proof -- this state may be viewed as implementing a deterministic checking automaton. The second is an adversarial liveness property, which requires its own proof system. The \Asp proof-checker takes a declarative proof sketch for an \Asp contract and turns that sketch into lemmas that encode the requirements of the proof system. The lemmas are checked by translation to an SMT solver, invoked indirectly through the Viper system~\cite{DBLP:conf/vmcai/0001SS16}. This provides \Asp users with deductive proofs of contract properties.

Deductive proofs are an important mechanism for  enhancing trust in the inherently trustless world of Web3, as any contract participant may \emph{independently} validate a claimed proof of a property against the contract code. As contract code is immutable once deployed, a proof remains valid throughout execution.

\section{\Asp By Example}
\label{sec:asp-by-example}

To illustrate how the design of \Asp promotes trustworthy smart contracts, we present an example of an open auction contract. Figure~\ref{fig:open-auction-skeleton} shows the skeleton of the contract written in \Asp. (We will fill in the contract gradually.)

\begin{figure}[ht]
\begin{lstlisting}[style=aspstyle, escapechar=`]
contract SimpleAuction(beneficiary: address, 
                       bidding_time: nat) 
  where beneficiary != Address.none && bidding_time > 0 {

  msg start, bid(coin);
  var tmr: timer, 
      maxBidder: address := Address.none;
    
  initial StartAuction;
    
  state StartAuction:
  | owner??start -> AuctionOpen `\label{line:by-owner}`
    { Timer.set(tmr, bidding_time); } 

  state AuctionOpen:
  | a??bid(c) `\label{line:a??bid}`
    when Timer.is_active(tmr) 
     &&  Coin.value(c) > Coin.value(maxBid) 
    notby beneficiary -> AuctionOpen `\label{line:notby}`
    { /* actions */ } 

  | when Timer.has_fired(tmr) -> AuctionClosed `\label{line:timer-has-fired}`
    { /* actions */ } 

  state AuctionClosed: // no transitions

}
\end{lstlisting}
\caption{Skeleton of an open auction contract in Asp}
\label{fig:open-auction-skeleton}
\end{figure}

\subsection{State Machine Structure}
\label{subsec:state-machine-structure}

An auction is naturally expressed as a state machine. The initial state is denoted by keyword \aspinline{initial}, the auction moves to a state where it is open and bidders can submit bids, and finally transitions to a state where the auction has closed. 
Mirroring blockchain execution, transitions can only be triggered by external messages (akin to function calls in function-based languages). The construct \aspinline{a??bid(c)} in line~\ref{line:a??bid} of Figure~\ref{fig:open-auction-skeleton} denotes a message named \aspinline{bid}, received from an address dubbed \aspinline{a}, containing an input parameter \aspinline{c} of type \aspinline{coin}.\footnote{Section~\ref{subsubsec: abstractions} elaborates on Asp's special abstract types.}

An \Asp contract starts in its initial state, and execution must follow the contract transition partial function $T: S \times M \rightharpoonup P(S)$, which maps each state and its input messages to the set of its possible next states.
Thus, if a contract is in state $s \in S$, a message $m \in M$ cannot be received at that state if $(s,m) \notin T$. 
(Contrast this with a function-structured language, where functions represent state \emph{changes} but there is no clear delineation of allowed state \emph{sequences}.) 

\Asp has transition guards to capture the conditions under which a transition is enabled. Transition guards include input guards denoting received messages, such as \aspinline{a??bid(c)} in line~\ref{line:a??bid}; predicates which must evaluate to true, such as \aspinline{when Timer.has_fired(tmr)} in line~\ref{line:timer-has-fired}; and access control, such as \aspinline{notby beneficiary} in line~\ref{line:notby}. 
Access control errors represent one of the most highly exploited contract vulnerabilities in 2023.
\footnote{\url{https://blog.merklescience.com/hubfs/Marketing\%20reports/Hackhub\%202024\%20(24-May-2024-7.29pm).pdf}}
\Asp's access guards make access control explicit. 

A transition without an input guard is called a $\tau$ transition. It is enabled at a state if its boolean guard evaluates to true. Although executable contracts must be deterministic, \Asp contracts may be internally non-deterministic -- that is, multiple $\tau$ transitions may be enabled at a state -- to allow for a notion of contract refinement. The compiler enforces determinism by choosing arbitrarily between enabled $\tau$-transitions; correctness proofs are not affected by this choice.

In Figure~\ref{fig:open-auction-snippet}, we add actions to the transitions from state \aspinline{OpenAuction} that specify state changes.
Transition actions are comprised of a \emph{loop-free} sequence of operations that include message transmissions and state updates. Restricting the action to be loop-free simplifies analysis, in effect by letting the skeleton structure explicitly define any loops.

\begin{figure}[h]
\begin{lstlisting}[style=aspstyle, escapechar=`]
contract SimpleAuction(...) {
  ...
  state AuctionOpen:
  | a??bid(c) `\label{line:receive-bid}`
    when Timer.is_active(tmr) 
      && Coin.value(c) > Coin.value(maxBid) 
    notby beneficiary -> AuctionOpen 
      { maxBidder!!bid_lost(maxBid); `\label{line:send_maxbid}`
        /* store new maxes */
        maxBidder = a;
        Coin.moveall(c,maxBid); } `\label{line:coin-moveall}`

  | when Timer.has_fired(tmr) -> AuctionClosed
    { beneficiary!!winner(maxBid, maxBidder); } 
  ...
}

contract Beneficiary() {
  var auction: Address;
  ...
  state AcceptBid:
  | a??winner(amt, addr) by auction -> FinalState { `\label{line:receive-amt}`
    log!!final_winner(Coin.value(amt)); `\label{line:log_amt}`  
  }
  ...
}
\end{lstlisting}
\caption{Code snippets of open auction contract and receiving contract in Asp}
\label{fig:open-auction-snippet}
\end{figure}

\subsection{Abstractions}\label{subsubsec: abstractions}

We introduce the \emph{abstractions} of common Web3 constructs that are included in \Asp. We demonstrate how these abstractions simplify verification by comparing the \Asp types and semantics with existing encodings and verification schemes.
\subsubsection*{Basic Types} Basic types include \verb|int| (integer), \verb|nat| (naturals), \verb|Tuple| (tuples), \verb|Seq| (unbounded sequences), and \verb|Map| (mappings from a key type to a value type).

In a significant departure from a conventional programming language, the number types (\verb|int| and \verb|nat|) are given their \emph{mathematical} definitions. As a consequence, there is no notion of arithmetic overflow in \Asp, so a contract programmer need not be concerned with vulnerabilities that arise from such overflow. 
Of course this is a fiction. It is the task of the defensive compiler to check for violations of this fiction (such as when arithmetic operations overflow) and cancel the transaction execution when violations occur. Certain operations are inherently partial: for instance, division by zero is undefined, as is an out-of-bounds access  to a sequence. The defensive compiler also checks that every executed operation is well-defined. Section~\ref{sec: compilation} elaborates.

\subsubsection*{Coins}
In conventional contract languages, there are two separate ways of accounting for cryptocurrency. The underlying blockchain maintains an account balance for each address and ensures that no double-spending can occur. At the contract level, cryptocurrency is represented as integers. This leaves contract-level accounting open to arithmetic error. Mistakes at the contract-accounting level cannot propagate to the blockchain  balances, but they can 
allow malicious contracts to obscure the amounts of currency sent or received by a contract.

\Asp unifies both views in a single \aspinline{Coin} datatype. A \aspinline{coin} variable is a container for the native cryptocurrency of the blockchain in its most basic unit, whatever that may be. This allows \Asp contracts to be ported to multiple blockchains, each with its own cryptocurrency. Coins cannot be created: they may only be transferred. To enforce this requirement, the Coin datatype provides only three operations: \verb|Coin.value(c)| is the (non-negative integer) value contained in \verb|c|; \verb|Coin.moveall(c,d)| transfers all value from \verb|c| to \verb|d|; and the partial operation \verb|Coin.move(c,k,d)| transfers value $k$ from \verb|c| to \verb|d|, if \verb|c| has value at least $k$--if not, the operation is undefined. 
\begin{example}
2vyper~\cite{DBLP:journals/pacmpl/BramEMSS21} provides \emph{resources} as special ghost state that can only be manipulated in certain prescribed ways -- somewhat similar to the \aspinline{coin} type in \Asp. However, 2vyper users must define ghost state operations and set up coupling invariants to prove that the values of \verb|uint| type used for cryptocurrency accounting remain consistent with the resource values in ghost state.
\footnote{See, for example, Figure 9 on page 16 of~\cite{DBLP:journals/pacmpl/BramEMSS21}.}
\Asp users rely on the guarantees of \aspinline{coin} type, without needing extra proofs to verify against malicious accounting manipulation.
\end{example}

All coins that are received in an input message must be transferred to coin state variables--i.e., received coins are not lost. (This is enforced by the defensive compiler.) Sending a message containing coin variables transfers their entire value  to the receiving entity; thus, those variables have zero value after the send action.
\begin{example}
Line~\ref{line:coin-moveall} of Figure~\ref{fig:open-auction-snippet} moves all coins sent 
as a bid to the contract's \verb|maxBid| coin. Line~\ref{line:send_maxbid} first transfers the previous \verb|maxBid| to the dethroned \verb|maxBidder|, so that the final value of \verb|maxBid| is that of the accepted bid. 
\end{example}

The \verb|Coin| operations in \Asp directly modify the value of a coin operand. However, coins stored in \verb|Map|, \verb|Seq|, or \verb|Tuple| types are \emph{copied} when retrieved from their storing structure. To prevent spurious coin creation or destruction, coin operators only accept operands that \emph{reference} coins stored in complex types. \verb|Map.ref| (and likewise for \verb|Seq| and \verb|Tuple|) directly modifies the specified entry, so that total coins are conserved.

\begin{example}
    In Example~\ref{ex:token-transfer}, a map stores \verb|Token| types (described next; analogous operations to \verb|Coin|). Note that \aspinline{Token.move} expects \aspinline{Map.ref} rather than \aspinline{Map.get}, as \verb|move| operations  modify the values of their operands.
\end{example}

The \aspinline{coin} operations enforce a \emph{coin conservation} law: coins are neither created nor destroyed and all coins in circulation are accounted for. In Solidity contracts, one would have to explicitly prove that the integer-based accounting in contracts matches the blockchain-based accounting; the \Asp \aspinline{coin} datatype eliminates the need for such proofs.

\subsubsection*{Tokens}
While cryptocurrency conservation is maintained by the blockchain, tokens in conventional languages have no such protection and are exclusively treated as \verb|uint|s. Issuing, burning, and transferring tokens is entirely comprised of arithmetic manipulation, heightening the need to verify against malicious token contracts.
\Asp provides an abstract \aspinline{token} type to prevent honeypot contracts and other exit scams that issue malicious tokens.\footnote{\url{https://certik.com/resources/blog/honeypot-scams}}

A \emph{token} in \Asp is essentially a coin, with two important differences: (1) tokens are defined and allocated by an issuer, and (2) tokens can be of different kinds. In \Asp, these aspects are handled by designating an \Asp contract as a token issuer with the declaration ``\lstinline[style=aspstyle]{issues Token(<limit>)},'' where the optional \lstinline[style=aspstyle]{<limit>} is a natural number indicating the number of available tokens. 

A token issuing contract issues just one kind of token. Tokens are issued into a token variable \aspinline{v} through \aspinline{Token.issue(<number>,v)}, which  is defined only if a sufficient quantity of tokens are available for issue. The coin operations have analogous token operations 
(\aspinline{Token.move(v,k,w)}, \aspinline{Token.moveall(v,w)}, and \aspinline{Token.value(v)}) 
and follow similar token-conservation rules, up to minting (i.e., issuing) and burning (i.e., removal) through a \aspinline{Token.burn(v,k)} operation. 

\begin{example}
The following code adapted from Certik\footnote{\url{https://certik.com/resources/blog/honeypot-scams}} presents a snippet from malicious token contract:
\begin{lstlisting}[style=soliditystyle]
function setBalance(address user, uint256 amt) public onlyRole(DEFAULTADMINROLE){
  _balanceaccs[user] = amt * 10**decimals();}
\end{lstlisting}
This code allows the contract owner to change a user's token balance to any amount they specify. Such a scam is impossible with \Asp's \aspinline{token} type.
\end{example}

\begin{example}
\label{ex:token-transfer}
The Move~\cite{DBLP:conf/tacas/DillGPQXZ22} language is designed for contract verification. However, Move does not have a built-in notion of tokens: \cite{DBLP:conf/tacas/DillGPQXZ22} provides an example contract for designing a safe token.\footnote{Figures 1,2 on pages 3,4 in~\cite{DBLP:conf/tacas/DillGPQXZ22}.} 
To ensure that tokens are not generated or destroyed spuriously in a token transfer, the Move contract defines a transfer function as a withdraw followed by a deposit, both carefully defined to disallow manipulation. \Asp condenses this into a few lines of code with the \aspinline{token} type:
\begin{lstlisting}[style=aspstyle]
contract BasicCoin() issues Token() {
  msg transfer(nat,address); 
  var accounts: map[address,token]; 
  ...
  state Bank:
  | a??transfer(x,b) when Map.in(a,accounts) && Map.in(b,accounts) notby b -> Bank 
    {Token.move(Map.ref(accounts,a),x,Map.ref(accounts,b));}
}
\end{lstlisting}
The Move contract ensures that tokens are not duplicated by defining a unique token storage struct for each address. \Asp tokens cannot be duplicated by design. As tokens are explicit types, the \Asp contract simply has an \aspinline{accounts} map from \aspinline{address} to \aspinline{token}, through which tokens may be issued, transferred, or burned.
\end{example}

\subsubsection*{Addresses}
An \aspinline{address} holds the blockchain address of an \Asp contract or an external party. A contract contains the special address variables \aspinline{creator} and \aspinline{owner}. These are set at the time of instantiation to the address of the creating entity. The \aspinline{owner} may be modified, but only through an \aspinline{Address.change_owner} action initiated by the current owner. The \aspinline{creator}  cannot be modified. A contract instance has its own unique constant address, referred to as \aspinline{Address.self}.
\begin{example}
Line~\ref{line:by-owner} of Figure~\ref{fig:open-auction-skeleton} specifies that the first transition of the auction contract can only be triggered by the contract owner.
\end{example}

Addresses may be stored, copied, and transferred between contracts. Message receive and send operations refer to addresses. 
A special address, \aspinline{log}, is used to log messages to the blockchain. Log transfers are always enabled and do not change the state of the contract. 
\begin{example}
    Line~\ref{line:log_amt} of Figure~\ref{fig:open-auction-snippet} logs the amount of the winning bid after the auction is complete.
    \footnote{Interaction between contracts will be addressed in Section~\ref{subsec:interacting-contracts}.}
    This is akin to an event in Solidity.
\end{example}

\subsubsection*{Timers}
In blockchains, time is not measured as real time, but rather by the growth of the chain, to ensure that all miners and validators have a common view of time. As a consequence, the progress of time may be uneven and is not guaranteed, which may be a point of fallacy for developers. Time-dependent actions are important for contracts such as auctions, which must terminate, and for contracts such as hashed timed locks, which place a limit on how long cryptocurrency is kept in escrow.

\Asp includes a \aspinline{Timer} data type. A \aspinline{timer} variable represents a timer that is initially inactive. Timer operations move the (implicit) timer state machine through a sequence of states: \aspinline{Off}, \aspinline{Active(k)} (for $k>0$), and \aspinline{Fired}. The  \aspinline{Timer.has_fired} predicate determines whether a timer is in its \aspinline{Fired} state.  All timers advance together by the same non-deterministically chosen amount on a transition, which models the way time advances on the underlying blockchain. 
Timers also simplify proofs of liveness properties; Section~\ref{subsec:liveness} elaborates.
\begin{example}
VerX~\cite{verx}, a contract verification tool, provides a benchmark of a contract that runs a continuous sale, which divides time into ``buckets'' of twelve hours (in Solidity) for accounting purposes.\footnote{\url{https://github.com/eth-sri/verx-benchmarks/blob/master/Mana/main.sol}}
Since time varies based on the frequency of blocks added to the chain, the only guarantee that this contract provides is that a bucket will eventually be reset. However, it does not guarantee anything more precise about the time of each bucket, since time is dependent on the length of the blockchain.
\Asp timers abstract away concrete notions of time so users do not rely on mistaken assumptions. The snippet below replaces the concept of ``hours'' with a timer that progresses non-deterministically and is reset upon firing.
\begin{lstlisting}[style=aspstyle]
state TrackSale:
  | when Timer.is_active(timer_bucket) -> CheckMax 
    { ... }

  | when Timer.has_fired(timer_bucket) -> CheckMax 
    { Timer.reset(timer_bucket);
      Timer.set(timer_bucket, bucket_size);
      ...
    }
\end{lstlisting}
\end{example}

\subsection{Interacting contracts}
\label{subsec:interacting-contracts}
In our auction example, we may want to create a separate state machine to describe the actions of the beneficiary, perhaps to model sending the prize or dividing the winning bid among collaborators. Interactions between contracts are specified by synchronized send and receive actions, using notation akin to CCS and CSP. In Figure~\ref{fig:open-auction-snippet}, the second transition from \aspinline{AuctionOpen} in the \aspinline{SimpleAuction} contract sends the message \aspinline{winner} with parameters of type \aspinline{coin} and \aspinline{address} to the beneficiary. 
\footnote{The lack of withdraw pattern may seem concerning to smart contract developers who are familiar with reentrancy. Asp contracts automatically lock against reentrancy, so refunds can be sent to losing bidders directly.}
The contract \aspinline{Beneficiary} receives a corresponding transition message \aspinline{winner} of identical parameter types.
When \aspinline{SimpleAuction} sends the \aspinline{winner} message to \aspinline{Beneficiary}, the message will be received by \aspinline{Beneficiary} only if \aspinline{Beneficiary} previously set the variable \aspinline{auction} to the address of \aspinline{SimpleAuction}, and is at state \aspinline{AcceptBid}.

\section{\Asp Semantics}

At its core, \Asp contracts induce reactive state machines which communicate through synchronized message exchanges. We define the semantics of execution and communication. 
We begin by defining the transition semantics of a single contract instance, then consider the semantics of interactions between multiple contract instances. 

For a simpler notation, we assume that every contract transition contains either an input guard and no output actions, or is a $\tau$ transition with  at most one output action. A contract is easily restructured to meet these requirements by introducing fresh states and edges. 

\subsection{Single Instance Transition Semantics}\label{sec:semantics}

A single contract instance defines a labeled transition system with transitions labeled as input, output, or internal. The contract skeleton may be viewed  as the tuple $(Q,q_0,\delta)$, where $Q$ is the finite set of states, $q_0$ is the initial state, and $\delta \subseteq Q \times Q$ is the next-state relation. We refer to a pair $(q,q')$ in the next-state relation as an \emph{edge}. The state variables $V$ induce the  space of assignments $X = V \rightarrow D$. (For simplicity in notation, all variables have domain $D$.) The initial assignment is denoted $x_0$.

Every message declaration $m(t_0,\ldots,t_{n-1})$ is viewed semantically as the set of input letters of the form $m(\alpha,d_0,\ldots,d_{n-1})$, where $\alpha$ represents the address of the message sender and each $d_i$ is a value in $D$. The set of input letters is denoted $\Sigma$; this has a corresponding set $\out{\Sigma}$ of output letters. For a letter $e$ (input or output), $\out{e}$ represents the matching letter (output or input, resp.), with $\out{\out{e}} = e$. 

The labeled transition system for a single contract instance is defined as a tuple $(S,s_0,\Sigma,T)$. The state space $S$ is $Q \times X$. The initial state $s_0$ is $(q_0,x_0)$. $T$ denotes the set of labeled transitions. A transition from state $(q,x)$ to state $(q',x')$ is defined if $(q,q')$ is an edge labeled with guard $g$ and action $a$, the predicates in guard $g$ evaluate to true at the state $(q,x)$, the operations in action $a$ are fully defined at state $x$, and $x'$ is the result of performing the operations in $a$. This transition is labeled by input letter $e$ if $g$ has an input guard that evaluates to $e$ in the state $(q,x)$, by output letter $\out{e}$ if the action $a$ is an output action that evaluates to $e$, and by $\tau$ otherwise.

\subsection{Multiple Instance Semantics}
\label{sec:mult-inst-sem}
Consider a collection $M_1,\ldots,M_n$ of contract instances. Intuitively, the instances communicate by synchronizing pairwise on transition labels, i.e., when an output transition of one contract matches with an input transition of another. 

As described previously, contract execution on a blockchain is single-threaded and externally triggered. To match this execution model, we define a single-threaded \emph{cascading semantics} for \Asp contracts. Intuitively, a cascade is started at a quiescent configuration by invoking an input transition in one of the contracts. (A configuration is a vector of contract states; it is quiescent if no cascade is in progress.) This input transition may trigger a synchronized transition with another contract; the execution thread then moves to that contract. A cascade continues in this manner until no further synchronizations are possible.

The precise formulation of a cascade relies on a pushdown stack of contract indexes and is parameterized by a recurrence limit $R\geq 0$. A \emph{configuration} is a pair $(s,\gamma)$ where $s$ is a vector with $s(i)$ being the state of $M_i$, for all i, and $\gamma$ is a pushdown stack with entries from $\{1,\ldots,n\}$. The stack is represented as a sequence with the left end of the sequence being the top of the stack. It is an invariant of the semantics that in any reachable configuration the stack contains at most $R+1$ occurrences of each index. A \emph{quiescent} configuration is a pair $(s,\gamma)$ where the stack $\gamma$ is empty, denoted by $\epsilon$. We say that an input or output letter $e$ is \emph{directed towards the environment} if its address entry is not one of the $M$-contract instances. We use the update notation $s'=s[k \becomes u]$ to represent the state vector $s'$ which is identical to $s$ except at the $k$'th entry, which is $u$.

The transitions from a configuration $(s,\gamma)$ are as follows. 
\begin{enumerate}
\item (Local $\tau$-Move) If $k$ is the entry at the top of the stack and $(s(k),\tau,t)$ is a transition of $M_k$, then $((s,\gamma),\tau,(s',\gamma))$ is a transition, where $s' = s[k \becomes t]$. 

\item (Synchronized Push) If $k$ is the entry at the top of the stack and there is an output transition $(s(k),\out{e},t)$ in $M_k$ and a matching input transition $(s(l),e,u)$ in $M_l$ for some $l$ \emph{that has at most} $R$ occurrences in $\gamma$, then $((s,\gamma),\tau,(s',\gamma'))$ is a synchronized transition, where $s' = s[k \becomes t, l \becomes u]$ is the new state vector, and $\gamma' = l\gamma$ is the new stack.

\item (Environment Output) If $k$ is the entry at the top of the stack and there is an output transition $(s(k),\out{e},t)$ in $M_k$ where $e$ is directed towards the environment then $((s,\gamma),\out{e},(s',\gamma))$ is a transition, with $s' = s[k \becomes t]$. 

\item (Pop) If $k$ is at the top of the stack and none of the above types of transitions are enabled at $s(k)$, then $((s,\gamma),\tau,(s,\gamma'))$ is a transition, where $\gamma=k\gamma'$.

\item (Environment Input) Consider a quiescent configuration $(s,\epsilon)$. If there is an input transition $(s(k),e,u)$ in $M_k$ for some $k$ where $e$ is directed towards the environment, then $((s,\epsilon),e,(s',\gamma'))$ is a transition, where $s' = s[k \becomes t]$ is the new state vector, and $\gamma' = k$ is the new, non-empty stack.
  
\end{enumerate}

Along a computation, a \emph{cascade} is an execution fragment that starts at a quiescent configuration and ends at the next quiescent configuration. Every infinite computation can be partitioned into either an infinite sequence of cascades, or into a finite sequence of cascades followed by an infinite suffix where the stack is ``stuck'' and every transition is either a local move or an environment output.

Reentrancy attacks are blocked with $R=1$ as it is impossible for an attacker contract to trigger an account withdrawal transition twice within a cascade. (Appendix~\ref{sec:cascade-reentrancy} contains an illustrative example.)

\section{\Asp Verification Proof Methods}

By design and semantics, \Asp contracts are inherently free from common vulnerabilities such as reentrancy, out of bounds accesses, and arithmetic overflow. Other security properties, such as proper access control, must be established through verification. In \Asp, a user-supplied deductive proof is required for every claimed property of an \Asp contract. This has a crucial benefit in the trustless Web3 setting, as any user can \emph{independently} validate claimed proofs before entering into a contract.

The \Asp proof-checker reads in a proof sketch for an \Asp contract. It  translates the sketch into Hoare triples (per contract transition) according to the proof rule specified. The triples are checked for validity through Viper~\cite{DBLP:conf/vmcai/0001SS16}, an intermediate verification language based on SMT solving. 

Most attacks on smart contracts, including many classified as security violations, target violations of safety properties. Termination is guaranteed in practice by contract execution's reliance on available ``gas,'' cryptocurrency paid to a validator in return for executing a contract transaction. We address other liveness concerns, such as reachability and lockouts. 
The \Asp proof-checker currently supports safety and reachability verification, which we illustrate with examples.
Appendix~\ref{app:proofs-adversaries} describes a lockout vulnerability, its resolution, and a formal proof of lockout-freedom; implementation of this proof method is in progress.

\subsection{Safety Proofs with Ghost Variables}
Informally, a safety property is one that can be falsified in finitely many steps. (A precise formulation is in~\cite{DBLP:journals/ipl/AlpernS85}.)

A standard proof method for safety uses a finite-word automaton to detect violations. Instead of defining automata over subsets of atomic state \emph{propositions}, as is typical, we define the automaton directly over the contract state space $S$. A deterministic finite-word automaton recognizing \emph{violations} of a safety property $L$ is a tuple $A= (Q,q_0,\delta,R)$ where $Q$ is the set of automaton states (not necessarily finite), $q_0$ is the initial state;  $\delta: Q \times S \to Q$ is the transition function, and $R$ is a subset of states, which we refer to as the rejecting states. 

A \emph{run} $\rho$ of the automaton on an infinite computation $w=s_0,e_0,s_1,e_1,\ldots$ is  a function from $\mathsf{Nat}$ to $Q$, where $\rho(0)=q_0$ and the tuple $(\rho(i),s_i,\rho(i+1))$ is in $\delta$ for all $i$. The run is rejecting if $\rho(k)$ is in $R$ for some $k$. Computation $w$ violates $L$ if there is a rejecting run of the automaton on $w$.

Following the automaton-theoretic view of verification, we define a product transition system $M \times A$ from the contract machine $M$ and the safety-violation automaton $A$. This has state space $S \times Q$, initial state $(s_0,q_0)$, and transitions of the form $((s,q),(s',q'))$ where for some $e$, $(s,e,s')$ is in $T$ and $\delta(q,s) = q'$. It is straightforward to show that every computation of $M$ is safe if, and only if, a reject state of $A$ is not reachable in $M\times A$; equivalently, if the property ``not in $R$'' is invariant over the transition system $M \times A$.

In \Asp, the product construction is carried out manually using auxiliary state variables declared with the prefix \aspinline{ghost}. The type checker ensures that ghost state does not influence normal contract execution: ghost state is never used to modify contract state, ghost state is never communicated to other contracts through messages, and assertions on ghost state cannot be used to control transition enabledness. Ghost variables encode the state of the violation automaton. As the automaton is deterministic, automaton state updates can be added to every contract transition. (Updates to ghost/automaton state can, and do, depend on the contract state variables.) 
Checking that ``not in $R$'' is invariant amounts to showing that an assertion $\theta$ on the joint state (contract + ghost) satisfies: 
\begin{itemize}
\item (Initiality) $\theta(s_0,q_0)$ holds,
\item (Inductiveness) If $\theta(s,q)$ holds and $((s,q),(s',q'))$ is a transition of $M\times A$, then $\theta(s',q')$ holds, and
\item (Sufficiency) If $\theta(s,q)$ holds, then $q$ is not in $R$.
\end{itemize}

A safety proof sketch in \Asp partitions $\theta$ across the contract skeleton states as a family of assertions $\{\theta_m\}$, where $\theta_m$ is associated with the skeleton state $m$, for all $m$. Given these assertions, the proof checker carries out the initiality, inductiveness, and sufficiency checks through a translation of \Asp constructs to the Viper verifier, as demonstrated in the following example.

\subsubsection{Illustrating \Asp safety proofs}
\label{subsec:illustrating-safety-proofs}
In our auction example, one property we may want to prove is that all bidders receive proper refunds. To prove this property (via invariance), we add ghost variables to the contract (as shown in Figure~\ref{fig:open-auction-ghost}) to keep track of coins bidded and refunded. 

\begin{figure}
\begin{lstlisting}[style=aspstyle]
contract SimpleAuction(...) ... {
  ...
  ghost var bidded: map[address, int] default 0, 
            refunded: map[address, int] default 0;

  state AuctionOpen:
    | a??bid(c) ... -> AuctionOpen 
      {Map.set(bidded,a,Map.get(bidded,a) + Coin.value(c));
       Map.set(refunded,maxBidder, 
         Map.get(refunded,maxBidder) + Coin.value(maxBid));
       maxBidder!!bid_lost(maxBid);
       ... }
    | when Timer.has_fired(tmr) -> AuctionClosed 
      {Map.set(refunded,beneficiary, 
         Map.get(refunded,beneficiary) + Coin.value(maxBid)
       );
       beneficiary!!winner(maxBid, maxBidder);}
    ...
} 
\end{lstlisting}
\caption{Open auction contract with ghost variables}
\label{fig:open-auction-ghost}
\end{figure}

\Asp users specify proof sketches in a separate proof file. 
To prove that all dethroned bidders are refunded their full bids, we write the following assertion using ghost variables:
\begin{lstlisting}[style=aspstyle]
always forall b: address : 
 (b != maxBidder && b != beneficiary) 
 ==> Map.get(refunded, b) == Map.get(bidded, b)
\end{lstlisting}

\aspinline{always} denotes an assertion that holds at every skeleton state. To prove that the winning bidder is refunded their previous losing bids, we add:
\footnote{Note that this second property requires case-specific assertions. This is because \aspinline{Coin.value(maxBid)} is $0$ after the coin is sent to the beneficiary.}
\begin{lstlisting}[style=aspstyle]
always Map.get(refunded, beneficiary) == 0 ==> 
 Map.get(refunded, maxBidder) == 
 Map.get(bidded, maxBidder) - Coin.value(maxBid)
always Map.get(refunded, beneficiary) > 0 ==> 
 Map.get(refunded, maxBidder) == 
 Map.get(bidded, maxBidder) - Map.get(refunded, beneficiary)
\end{lstlisting}

We then add that the highest bidder is never the beneficiary, and require the following skeleton-state-specific assertions to support the proof:

\begin{lstlisting}[style=aspstyle]
@StartAuction Map.get(refunded, beneficiary) == 0           
@AuctionOpen Map.get(refunded, beneficiary) == 0
\end{lstlisting}

The \Asp proof-checker checks the initiality of these assertions and their inductiveness over every contract transition. (Sufficiency is not needed, as the assertion encodes the desired property directly.) The inductiveness checks turn into Hoare triples, which are compiled to Viper and verified automatically.

\subsection{Proofs of Timer-supported  Reachability}

Timers in an \Asp contract can be used to check reachability. However, the state-changes of the corresponding timer state machines are implicit, as is the progress of time which, in a blockchain, is measured by the non-uniform measure of blockchain length. To incorporate these implicit progress measures, we add a self-loop time-progression transition, $\Atime$, to every skeleton state. The $\Atime$ transition is enabled at state $s$ only if there is some timer in an active state at $s$. Its effect is to advance time by at least one unit and update the state of all active timers accordingly. (The $\Atime$ transition guard ensures that it is never enabled for a contract without timers.)

With this addition, the reachability proof scheme is formally as follows. A proof consists of a state assertion $\theta$ and a partial rank function $\rho$ over a well-founded relation $\prec$ that meet the following conditions (where $R$ represents the set of states for which we wish to prove reachability): 
\begin{enumerate}
    \item $\theta$ holds of the initial state, 
	\item $\rho$ is defined for all states in $\theta$ that are not in $R$, and 
 \item  For every state $s$ in $\theta$ but not in $R$: 
 \begin{enumerate}
    \item Some transition (either an explicit contract  transition or the implicit $\Atime$ transition) is enabled at $s$, and 
    \item For every transition from $s$ to a state $t$, it is the case that either $R$ holds at $t$, or $\theta$ holds at $t$ and $\rho(t) \prec \rho(s)$
 \end{enumerate}
 \end{enumerate}

This proof method is sound for reachability. Consider, to the contrary, that there is a maximal contract computation that does not include a state in $R$. By the first and third rules, every state on this computation satisfies $\theta$ but not $R$. This computation cannot be finite, as the final state must have an enabled transition, contradicting maximality. Hence it must be infinite: but then it induces an infinite decreasing chain in $\prec$, which contradicts well-foundedness.

\subsubsection{Illustrating liveness proofs with timers}
\label{subsec:liveness}

We wish to prove that  \aspinline{AuctionClosed} will eventually be reached, to ensure that the auction will close and the beneficiary will receive the winnings.
The following timer-based proof verifies this.

\begin{lstlisting}[style=aspstyle]
reachability auction_closed(2) { // "2" is the rank length 
  goal = {
     @AuctionClosed true
     /* other state-specific goals are false by default */
  }
  invariant =  {
     @StartAuction  Timer.is_off(tmr)    
     @AuctionOpen   !Timer.is_off(tmr)
  }
  rank = { /* partial function, defined by cases */
     @StartAuction 
       | (2,0)  
     @AuctionOpen  /* Order is important */                
       | (1, 0) if Timer.has_fired (tmr) 
       | (1, Timer.value(tmr)) if Timer.is_active (tmr) 
     @AuctionClosed
       | (0, 0)
  }
  witness = {
       @StartAuction a==owner && a != Address.none
       @AuctionOpen  a != beneficiary && a != Address.none && Coin.value(c) > Coin.value(maxBid)
  }
}
\end{lstlisting}

The \Asp verifier compiles this proof outline to a set of lemmas for the Viper verification tool, which encode all the checks defined for the timer-based reachability proof rule. The enabledness check (3(a)) requires existentially quantifying the free variables of each transition. 
As existential quantification is not well-supported by Viper\footnote{From \url{http://viper.ethz.ch/tutorial/\#expressions}}, 
we explicitly write the existential witness necessary to show that transitions are enabled at each state.
The well-founded set is the set of natural-number tuples (of a fixed length), ordered lexicographically. Verification of the generated  lemmas takes less than $2$ seconds.

\subsection{Proofs of Liveness in Adversarial Environments} 

A class of attacks on contracts consists not in stealing cryptocurrency, but rather in ``freezing up" the contract so that it becomes unusable, so that any funds stored in the contract are inaccessible. Showing that such lockouts are not possible requires reasoning about potential adversarial actions in the multi-agent setting of Web3.  

One way to formulate the property is to do so in game-theoretic terms. We consider an external Player (address) $x$ and show that from any reachable contract state, there is a winning strategy for the Player to reach a contract state satisfying property $Q$ (say a state where a message from $x$ must be accepted). The Opponents are the other agents and the contract $M$ itself, as $M$ contains non-deterministic $\tau$-actions which are resolved arbitrarily. 

The following deductive proof system establishes this property. It is inspired by similar proof systems (cf.~\cite{DBLP:conf/cav/Namjoshi01}) for $\mu$-calculus properties.
A proof consists of a state assertion $\theta$ and a partial rank function $\rho$ that meet the following conditions: 
\begin{enumerate}
    \item $\theta$ is an invariant of the contract $M$,
	\item $\rho$ is defined for all states in $\theta$, and 
 \item  For every state $s$ in $\theta$, one of the following holds: 
 \begin{enumerate}
    \item $s$ satisfies $Q$, or 
    \item There is a transition for the Player to a state $s'$ that is in $\theta$, and the rank decreases strictly after that transition, or
	\item Some Opponent transition is enabled and all Opponent transitions lead to states in $\theta$ and strictly decrease rank. 
 \end{enumerate}
 \end{enumerate}

The soundness of this proof rule is established as follows. Consider any reachable state, say $s$. As this state is reachable, it is in the invariant $\theta$. From this state, the choices of the Player and Opponent produce a game subtree where all tree states are in $\theta$. This tree cannot have an infinite branch where $Q$ never holds, as rank decreases strictly on every transition, but the domain of $\rho$ is well-founded. Thus, every branch must end in a state satisfying $Q$. The proof rule is also relatively complete: in fact, it is deduced from a $\mu$-calculus framing of the property (Appendix~\ref{app:proofs-adversaries}). 
Implementation of this proof system is in progress.

\section{Defensive compilation}
\label{sec: compilation}

The \Asp compiler translates \Asp contracts to Solidity, the most popular contract language; a translation to Ink! (and Rust) is in progress. To make the discussion concrete, we focus on Solidity and Ethereum; the translation to other languages and blockchains is similar. 

We briefly summarize the aspects of contract execution on Ethereum that are most relevant to compilation. On a blockchain, a  contract is passive until one of its interface methods is invoked by an external entity; this invocation is called a \emph{transaction}.  The externally-invoked method may recursively invoke other methods, including those of other contracts. A transaction is executed by the EVM (Ethereum Virtual Machine) in a single-threaded manner until completion. On successful completion, changes to the contract state are committed to the blockchain. On failure, which can be due either to an undefined instruction (e.g., divide by zero) or a programmed `revert' instruction, the state of the blockchain is not changed. Every transaction execution has a cost in cryptocurrency, known as the `gas' fee.

The compiler must transform the state-machine view of an \Asp programmer to the method-invocation view of the underlying contract execution engine. The compiler does so by essentially turning an \Asp state machine skeleton inside-out. Every \Asp message is transformed into a publicly accessible contract method. Within this method, a case analysis by (skeleton) state determines the transition that is executed and the next skeleton state. If no transition is enabled, the compiled contract reverts. Cascade semantics is implemented through a (private) \verb|tau_closure| method that repeatedly executes $\tau$ transitions from the next skeleton state on, until a state without $\tau$ transitions is reached. The source \Asp contract may be internally non-deterministic; the compiler determinizes execution by making an arbitrary choice between simultaneously enabled $\tau$ transitions. Message sends are converted to a low-level Solidity \verb|call| operation. Ghost variables and ghost operations are not compiled to Solidity; they are used only for proofs.

In addition, the compiler inserts code that, at run time, performs two functions: (a) it ensures that the assumptions underlying verification (such as no arithmetic overflow, no reentrancy, no undefined actions) hold during execution and (b) it checks for properties that may be cumbersome to prove, such as coin conservation. These checks guard against semantics violations and are thus defensive in nature; hence, we refer to the process as \emph{defensive compilation}. The no-reentrancy property is enforced by introducing a ``reentrancy counter'' that is checked then incremented on entry into every invocable method and decremented on exit. A reentrant call exceeding the reentrancy limit will fail the check and cause the entire transaction to be canceled.

The \Asp semantics has no notion of explicit failure. However, a transaction of the compiled Solidity code may be canceled due to the failure of a compiled \Asp guard, or due to the failure of an inserted defensive check, or due to a detected arithmetic overflow. The first two cases match the \Asp semantics, as the corresponding transition is not defined in \Asp. The third case is one where the \Asp transition is defined (as arithmetic in \Asp is ideal) but the compiled version does not succeed. As a consequence, compilation correctness is expressed not as an equivalence but as a language inclusion. This gives us the following important property for \Asp compilation.

\begin{property}\label{thm:compilation}
  Let $C_1,\ldots,C_n$ be \Asp contracts with corresponding compiled contracts $S_1,\ldots,S_n$. Every successful transaction of the compiled contracts corresponds to a successful cascade of the \Asp contracts. 
\end{property}

As a corollary, every safety property of an \Asp contract is a safety property of the compiled contract.
The corollary is important as it ensures that proof effort is required only at the \Asp level; assuming correct compilation, there is no need to re-check the code generated for (multiple) target blockchains. 

These correspondences also hold for adversarial liveness and reachability properties under the assumption that arithmetic overflow or shortage of gas do not cause a transaction to revert. We then obtain in the other direction that every \Asp contract cascade is matched by a successful transaction of the compiled code. This induces a bisimulation between the \Asp and compiled Solidity transition systems that preserves adversarial liveness and reachability.

\section{Related Work and Discussion}

The importance of ensuring that smart contracts are free of bugs was recognized early. In 2016, only a year after Ethereum was created, a hacker stole a large amount of ETH (worth about $\$3$B today) in the ``The DAO'' exploit. That led to a controversial decision to ``hard fork'' the Ethereum chain to recover the funds. This recourse is impossible today: there are far too many active (and buggy) contracts to hard-fork the chain on each exploit. 

Several companies (e.g., CertiK, Certora) offer smart contract audit services, which look for errors in smart contracts.\footnote{\url{http://www.certik.com} and \url{http://www.certora.com}.}  Tools such as MYTHRIL\footnote{\url{https://github.com/Consensys/mythril}} and the Solidity SMT tools~\cite{smt-solidity} search for errors using bounded model checking and automatic invariant inference. The tools and services are valuable for detecting potential errors, but they do not provide a comprehensive guarantee that a contract is free of vulnerabilities.  The VerX~\cite{verx} tool model-checks safety properties of Solidity contracts, while the SmartPulse~\cite{smartpulse} tool checks LTL properties. Both tools extensively apply predicate abstraction methods. Fully automated checkers are, however, inherently limited by state explosion. 

In Section~\ref{sec:introduction}, we point to the difficulty of formally proving security for contracts written in full-fledged languages such as Solidity. The solc-verify tool~\cite{solc-verify} checks user-supplied assertions on Solidity contracts, but it is restricted to quantifier-free assertions. Hence, one cannot typically write proofs about maps, for instance. 2vyper~\cite{DBLP:journals/pacmpl/BramEMSS21} is a verification system for the Vyper language, a variant of Solidity. As neither the Solidity nor the Vyper language restrict reentrancy, it is necessary to prove either that the contract code blocks reentrancy, or that reentrancy, if it occurs, does not lead to a vulnerability. The 2vyper system includes special proof rules to prove these assertions. While the proof system is technically interesting, such proofs add a substantial burden in practice. A similar reentrancy-sensitive proof system is defined in~\cite{DBLP:conf/fmics/CassezFQ22} for contracts written in the Dafny language (and potentially compiled to Solidity, although that is not yet implemented).

Such difficulties have led to the development of several non-standard programming languages for smart contract development (including \Asp) which eliminate reentrancy as an issue. Move~\cite{move-whitepaper-2019} is an object-oriented language with a non-standard, strict type system, similar to the ``borrow'' system of Rust. The type system naturally blocks reentrancy. The Move verifier~\cite{DBLP:conf/cav/ZhongCQGBPZBD20,DBLP:conf/tacas/DillGPQXZ22} checks user-supplied proofs and is impressive in its scope and application. However, the non-standard type system of the Move language must be enforced by the underlying virtual machine execution, which limits portability. Indeed, Move is tightly linked to the Diem blockchain and variants such as Aptos and Sui, and has not been ported to common blockchains such as Ethereum and Solana. 

Obsidian~\cite{obsidian} is another non-standard object-oriented contract language that uses typestate and linear typing to enable static verification of resource use. 
However, Obsidian does not in itself support the verification of user-supplied safety or liveness properties. 

While the languages discussed so far are conventional in their structure and constructs, FSolidM~\cite{DBLP:conf/fc/MavridouL18} models a contract as a finite-state machine skeleton that is extended with Solidity state variables and statements. The associated VeriSolid~\cite{DBLP:conf/fc/MavridouLSD19} tool checks CTL properties of sequences of transition labels of FSolidM machines, using automated data abstraction and model checking via BIP and the NuSMV tool.  
While model checking has its advantages, the scope of verified properties is restricted to propositional CTL (thus no arithmetic or quantified assertions), and the language is closely tied to Solidity.

Contracts in Scilla~\cite{DBLP:journals/pacmpl/SergeyNJ0TH19} are also represented as state machines. However, Scilla is viewed as an intermediate notation, with contracts compiled to Coq for verification. The Tezos blockchain also translates from its bytecode notation to Coq~\cite{DBLP:conf/fm/BernardoCHPT19}. The KEVM formalization of Solidity semantics~\cite{DBLP:conf/csfw/HildenbrandtSRZ18} is similar in nature. While translations to Coq permit complex properties to be expressed and verified, they also create a substantial proof burden, requiring considerable expertise and manual effort.  

With \Asp, we explore a different point in the design space. The \Asp language inherently blocks several commonly exploited vulnerabilities, eliminating a significant concern for programmers. Like the FSolidM designers, we believe that an explicit state-machine skeleton simplifies contract design. However, \Asp contracts do not operate on Solidity data types or expressions. Instead, \Asp contracts operate on abstract data types, whose mathematical definitions simplify the writing of contracts, and their analysis. Abstract operations are concretized by the compiler and are therefore correct by construction. 

Indeed, its reliance on abstract data types is one of the major distinguishing features of \Asp. A further advantage of programming with abstract types is portability: a verified \Asp contract can be compiled to multiple languages and blockchains. The \Asp type system is otherwise conventional (unlike that of Move), which faciliates compilation to standard languages such as Solidity, Rust, and Ink! and to commonly used blockchains such as Ethereum and Solana. 

\Asp, like Move, relies on programmer-supplied deductive proofs. We believe that this is crucial for security, as deductive proofs allow greater expressiveness (e.g., quantifiers to express properties of maps) and flexibility (e.g., safety, liveness, and adversarial proofs). A current drawback is that proof sketches for even simple properties must be supplied manually; in the future, we aim to augment programmer-supplied proofs with \emph{proof-generating} model checking~\cite{DBLP:conf/cav/Namjoshi01,DBLP:conf/fsttcs/PeledPZ01}. As argued previously, we believe that it is important in the context of Web3 to provide explicit, independently-checkable proofs for every security claim. In this sense, the design of \Asp follows the proof-carrying-code principle~\cite{DBLP:conf/popl/Necula97} of placing the burden of constructing a proof on the contract creator (who may use automated methods) while making it possible for every user to independently validate a claimed proof. 

Our motivation in developing \Asp is to explore the language design space, prioritizing ease of reasoning over sophisticated language features.
Abstractions compensate for the simpler language structure; the defensive compiler enforces the abstract semantics and strengthens security; and verifiable proofs build trust. 
We expect \Asp to evolve over time.\footnote{We intend to release the \Asp system as open source once we have the necessary approvals. At present, we have made several \Asp contracts (with proofs) available at \url{https://github.com/DebraChait/Asp-example-contracts}.} We are considering the introduction of new abstractions, such as commitments, secret inputs, and randomness, and modular proof methods. We also plan to explore how verified proof assertions can be used by the compiler to eliminate run-time defensive checks (cf. ~\cite{prisc-gradual-compilation-2024}).

\bibliographystyle{splncs04}
\bibliography{references}

\newpage
\appendix
\appendix
\section{Cascade semantics and reentrancy}\label{sec:cascade-reentrancy}
The cascade semantics of \Asp contracts supports limited reentrancy to allow patterns of sending a message and receiving a response, and to simultaneously prevent malicious reentrancy often characterized by the ability to reenter a contract arbitrarily many times.

To demonstrate the cascade semantics and the prevention of malicious reentrancy, we take a simple example of reentrancy from Solidity By Example.\footnote{https://solidity-by-example.org/hacks/re-entrancy/}
This involves two contracts, one which stores and returns ether (the cryptocurrency of the Ethereum blockchain), and one which attempts to drain all ether from the first contract via reentrancy. The two contracts are written as \Asp state machines in Figure~\ref{fig:reentrancy-example}, with the shorthand \aspinline{bal[a]} to represent \aspinline{Map.set} and \aspinline{Map.get} operations. We set the reentrancy limit $R = 1$.

According to \Asp's semantics for multiple interacting contracts (Section~\ref{sec:mult-inst-sem}, a cascade can only begin from a quiescent state with environment input, mirroring smart contracts' reliance on external calls to begin execution. We start our cascade with an Environment Input configuration transition via the Attacker contract's receipt of a coin through a "send" message. The cascade proceeds as follows (we note which configuration transitions are taken at each step):
\begin{enumerate}
    \item \emph{Environment Input, via Attacker}. Attacker is pushed onto the stack, and Attacker's state is updated to CollectDeposit.
	\item \emph{Synchronized Push}. Etherstore is pushed onto the stack. Attacker's state is updated to Etherstore Deposit, and Etherstore's state is updated to AcceptDeposit.
    \item \emph{Pop}. Attacker is now at the top of the stack.
    \item \emph{Synchronized Push}. Etherstore is pushed onto the stack. Attacker's state is updated to EtherstoreWithdraw, and Etherstore's state is updated to WithdrawRequested.
    \item \emph{Synchronized Push}. Attacker is pushed onto the stack (with Attacker at the bottom of the stack as well). Since $R=1$, this is allowed. Attacker's state is updated to AcceptReturn, and Etherstore's state is updated to ResetBalance.
    \item \emph{Pop}. Note that this is where Attacker attempts to maliciously reenter Etherstore. This reentrancy is blocked, because the limit of $R=1$ is exceeded. Attacker is popped off the stack.
    \item \emph{Local $\tau$-Move, followed by another Local $\tau$-Move}. Etherstore's state is updated to GaveWithdrawal, and then to AcceptDeposit.
    \item \emph{Pop}. Etherstore is popped off the stack.
    \item \emph{Pop}. Attacker is prevented from requesting another withdrawal, since its balance has already been reset. Attacker is therefore popped off the stack, and we return to a quiescent configuration.
\end{enumerate}

The reentrancy limit allows for an exchange of message and response, but prevents the unlimited reentrancy that characterizes malicious attacks. The cascade semantics then demand that the would-be-victim contract completes its state changes before the attacker can synchronize with it again.

\begin{figure}
\centering
\begin{subfigure}{0.9\textwidth}
\centering
\begin{lstlisting}[style=aspstyle]
contract Etherstore{
...
state AcceptDeposit:
| a??deposit(value) -> AcceptDeposit {bal[a] += value}
| a??withdraw when bal[a] > 0 -> WithdrawRequested {req = a}

state WithdrawRequested:
| -> ResetBalance {req!!return(bal[req])} 

state ResetBalance:
| -> GaveWithdrawal {bal[req] = 0}

state GaveWithdrawal:
| -> AcceptDeposit {}
}
\end{lstlisting}
\caption{Contract that stores and returns ether.}
\label{subfig:etherstore}
\end{subfigure}%
\vfill
\begin{subfigure}{0.9\textwidth}
\centering
\begin{lstlisting}[style=aspstyle]
contract Attacker{
...
state Start:
| c??send(amt) when amt >= 1 -> CollectDeposit {amount = amt}

state CollectDeposit:
| -> EtherstoreDeposit {Etherstore!!deposit(amount)}

state EtherstoreDeposit:
| -> EtherstoreWithdraw {Etherstore!!withdraw}

state EtherstoreWithdraw:
| b??return(balance) -> AcceptReturn {}

state AcceptReturn:
| -> Attack {if Etherstore.balance >= 1 then Etherstore!!withdraw}

state Attack:
| -> EtherstoreWithdraw {}
}
\end{lstlisting}
\caption{Contract that tries to drain Ether from the above contract via reentrancy.}
\label{subfig:attacker}
\end{subfigure}
\vfill
\caption{Two contracts that demonstrate limited reentrancy in \Asp's cascade semantics.}
\label{fig:reentrancy-example}
\end{figure}

%%%%%%%%%%%%%%%%%%%%%%%%%%%%%%%%%%%%%%%%%%%%%%%%%%%%%%%%%%%%%%%%%%%
\section{Vending Machines and Lockout-Freedom}
\label{app:proofs-adversaries}

\begin{figure}[ht]
\begin{lstlisting}[style=aspstyle]
contract VendingMachine {
  msg pay(coin), refund(coin);
  msg order, item, cancel, halt;
  
  var customer: address,
      paid, total: coin; 

  // states and state transitions
  initial Wait;

  state Wait: // for a new customer
   | x??pay(c) -> Choose 
       { customer = x; Coin.moveall(c,paid); }
   | owner??halt -> Halt

  state Choose: // place an order or cancel
   | customer??order -> Deliver { Coin.moveall(paid,total); }
   | customer??cancel -> Wait 
       { customer!!refund(paid); customer = Address.none; }

  state Deliver: // deliver item
   |  -> Wait { customer!!item; customer = Address.none; }

  state Halt: // halted, no actions
} 
\end{lstlisting}

\caption{A Vending Machine contract outline.}
\label{fig:simplevm}
\end{figure}

Figure~\ref{fig:simplevm} shows an outline of a vending machine, which suffices to illustrate lockouts. Consider the following desirable property: it should always be possible for any customer to use the machine--possibly after a short wait. 

This contract \textbf{does not} satisfy the property, for an interesting reason: an absent-minded -- or malicious -- customer may enter a coin, then simply  walk away, effectively locking the machine, which stays `frozen' in its \aspinline{Choose} state. 

To fix the problem, we may introduce a timer and a timeout at the Choose state. A different resolution is to allow \emph{any one}, not just the current customer, to cancel the current transaction at the \aspinline{Choose} state. Both options satisfy the critical safety property that the original customer is refunded their money on cancellation -- unlike in a real vending machine! 

This property is a form of eventual access or deadlock freedom ($\tA\tG\tE\tF$ in CTL) but in an adversarial, multi-agent setting. Lockout-freedom may be formulated as the property that \emph{every} actor $x$ has eventual access to the machine. Eventual access for actor $x$ is written as $\tA\tG\tE\{x\}\tF\,Q$, where by $\tE\{x\}F$ we mean that there is a winning strategy for the Player (actor $x$) that eventually reaches a state where $Q$ holds. The Opponents are the other actors in the environment of $M$, and $M$ itself, as the contract may make arbitrary $\tau$-transition choices.

The following deductive proof system establishes this property. It is inspired by similar proof systems (cf.~\cite{DBLP:conf/cav/Namjoshi01}) for $\mu$-calculus properties. A proof consists of a state assertion $\theta$ and a partial rank function $\rho$ that meet the following conditions: 
\begin{enumerate}
    \item $\theta$ is an invariant of the contract $M$,
	\item $\rho$ is defined for all states in $\theta$, and 
 \item  For every state $s$ in $\theta$, one of the following holds: 
 \begin{enumerate}
    \item $s$ satisfies $q$, or 
    \item There is a transition for the Player to a state $s'$ that is in $\theta$, and the rank decreases strictly after that transition, or
	\item Some Opponent transition is enabled and all Opponent transitions lead to states in $\theta$ and strictly decrease rank. 
 \end{enumerate}
 \end{enumerate}

The soundness of this proof rule is established as follows. Consider any reachable state, say $s$. As this state is reachable, it is in the invariant $\theta$. From this state, the choices of the Player and Opponent produce a game subtree where all tree states are in $\theta$. This tree cannot have an infinite branch along which $Q$ never holds, as that would induce an infinite strictly decreasing rank sequence, contradicting well-foundedness. Thus, every branch must satisfy $Q$. 

We show how to apply this proof method to the contract corrected by allowing any customer to cancel in the Order state. Here, $Q$ is the assertion that \aspinline{pay} is enabled for $x$; $\theta$ is essentially the set of all states, except that at the \aspinline{Choose} state there must be a defined \aspinline{customer}. The rank $\rho$ is defined on the naturals, with value $0$ at the \aspinline{Wait} state, value $2$ for the \aspinline{Choose} state with a defined customer, and value $1$ at the \aspinline{Deliver} state. Consider a customer $x$. If the machine is at the \aspinline{Wait} state, then $Q$ holds, as \aspinline{pay} is enabled for $x$. If the machine is at the \aspinline{Choose} state with a customer, say $y$ (possibly different from $x$), then the \aspinline{cancel} transition by $x$ (the Player) changes state to \aspinline{Wait} state, with a reduction in rank. There is no Player transition at the \aspinline{Deliver} state, but the $\tau$-transition (by the Opponent $M$) moves the contract to the \aspinline{Wait} state, with a reduction in rank. 

(The \Asp proof checker is currently being extended to support this and other proof rules for game-like properties.)

The (relative) completeness argument relies on a $\mu$-calculus framing of the property $\tE\{x\}Fq$, as $(\mu Y: p \lOr \tDia{\mathsf{\Player}}(Y) \lOr (\tDia{\Opponent}(\TT) \lAnd \tBox{\Opponent}(Y)))$. It essentially splits the least fixpoint according to stages, which induce the rank function -- the strategy used in~\cite{DBLP:conf/cav/Namjoshi01}.

\end{document}